\documentclass[
    ,final            % use final for the camera ready runs
  ]
  {aipproc}

\layoutstyle{6x9}

\newcommand{\minerva}[0]{MINERvA}
\newcommand{\nubar}[1]{\bar{\nu}_{#1}}

\begin{document}

\title{Quasi-Elastic Scattering in \minerva{}}


\author{Kevin S. McFarland}{
  address={Department of Physics and Astronomy, University of Rochester, Rochester, NY 14627 USA\\and Fermi National Accelerator Laboratory, Batavia, IL 60510 USA\\}
\vspace{4 mm}
\emph{On behalf of the MINERvA Collaboration}\footnote{http://minerva.fnal.gov}
}

\begin{abstract}
Determination of the quasi-elastic scattering cross-section over a
broad range of neutrino energies, nuclear targets and $Q^2$ is a 
primary goal of
the \minerva{} experiment.  We present preliminary
comparisons of data and simulation in a sample rich in
$\nubar{\mu}p\to \mu^+n$ events from approximately one
eighth of the total $\nubar{}$ events collected by \minerva{} to date.
We discuss future plans for quasi-elastic analyses in \minerva{}.
\end{abstract}

\maketitle

%%%%%%%%%%%%%%%%%%%%%%%%%%%%%%%%%%%%%%%%%%%%
%% MAINMATTER
%%%%%%%%%%%%%%%%%%%%%%%%%%%%%%%%%%%%%%%%%%%%

\begin{figure}
\mbox{
\includegraphics[width=.35\textwidth]{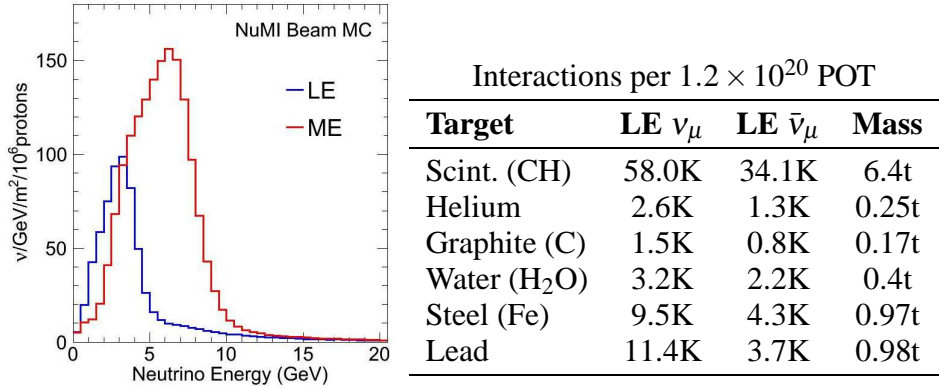} 
\raisebox{11ex}
{
\begin{tabular}{lccc}
\multicolumn{4}{c}{Interactions per $1.2\times10^{20}$ POT} \\
\hline
%  \tablehead{1}{r}{b}{Target}
%  & \tablehead{1}{r}{b}{LE $\nu_\mu$}
%  & \tablehead{1}{r}{b}{LE $\nu_\mu$}
%  & \tablehead{1}{r}{b}{Fiducial\\ Mass}   \\
  {\bf Target}
  & {\bf LE $\nu_\mu$}
  & {\bf LE $\bar{\nu}_\mu$}
  & {\bf Mass}   \\
\hline
Scint.\ (CH)& 58.0K & 34.1K & 6.4t\\ 
Helium & 2.6K & 1.3K & 0.25t\\ 
Graphite (C) & 1.5K & 0.8K & 0.17t\\ 
Water (H$_2$O) & 3.2K & 2.2K & 0.4t\\ 
Steel (Fe) & 9.5K & 4.3K & 0.97t\\
Lead & 11.4K & 3.7K & 0.98t\\
\hline
\end{tabular}
} % raisebox
} % mbox
\caption{(left)
  The Neutrino Flux in the NuMI Low Energy (LE) and Medium Energy (ME)
  beams.  Current \minerva{} data is from the LE configuration, and
  future running concurrent with the NOvA experiment will use a
  spectrum more similar to the ME beam.  (right) Fiducial CCQE
  Interactions in the Low Energy
  (LE) beam.  The fiducial mass is also given in metric tons for each
  target.}
  \label{fig:spectrumRates}
\end{figure}

The \minerva{} experiment was designed, in part, to study
charged-current quasi-elastic (CCQE) neutrino
($\nu_{\mu}n\to \mu^-p$) and anti-neutrino
($\nubar{\mu}p\to \mu^+n$) scattering of neutrinos with energy
$1$ to $10$ or more GeV over a broad range of $Q^2$.  \minerva{}
is currently running in the "low energy" (LE) NuMI beam designed for the
MINOS experiment in which it plans to integrate $4\times 10^{20}$
protons on target (POT)
in neutrino mode and $2\times 10^{20}$ POT in anti-neutrino
mode.  The anti-neutrino exposure is a mixture of data taken
with a partially constructed detector and the full \minerva{} detector.  
After a shutdown beginning in 2012, the NuMI beamline will
operate in the "medium energy" (ME) tune.  Figure~\ref{fig:spectrumRates}
shows the predicted neutrino flux in both the LE and ME beams.

The active elements of the \minerva{} detector~\cite{nuint11-schmitz} 
are polystyrene scintillator bars formed into detector planes.  A
continuous series of these planes forms the ``tracker'' region of
the \minerva{} detector.  The analysis reported in this paper used
events exclusively from the tracker; however, the \minerva{} detector
also contains a region where the planes of scintillator are
interspersed with passive nuclear targets.  These will allow study of
$A$-dependence in the CCQE rates at different energy or $Q^2$.
Figure~\ref{fig:spectrumRates} also shows the expected rate of
CCQE interactions in the scintillator tracker and in the passive
nuclear targets.  Of course, not all of these events will be accepted
in the analysis, but the combination of the significant mass of
the \minerva{} targets and the high intensity statistics available in
the NuMI beam are very promising for this measurement.

The planned results for CCQE in \minerva{} include measurements of the
CCQE total cross-section, $\sigma$, and $d\sigma/dQ^2$ on the
scintillator target.  At low to moderate $Q^2$,
$Q^2\stackrel{<}{\sim}1$~GeV$^2$, the primary goal is comparison with
results from K2K~\cite{K2K-CCQE-1,K2K-CCQE-2},
MiniBoone~\cite{MiniBooNE-CCQE-1,MiniBooNE-CCQE-2}, 
SciBooNE~\cite{SciBooNE-CCQE} and
NOMAD~\cite{NOMAD-CCQE}.  In particular, there are open questions
about the apparent cross-section enhancement at low energy and
moderate $Q^2$ and suppression at low $Q^2$ in the existing
data. \minerva{} will repeat these measurements on its multiple
nuclear targets to look for changes with increasing $A$ and nuclear
density.  At higher $Q^2$, \minerva{} will be able to make the first
measurements of the axial form factor and compare the high $Q^2$
behavior with vector form factors.  \minerva{} also plans to
compare $\nu_e$ CCQE with the same $\nu_\mu$ process using the 
$\sim 1.5\%$ of its events from $\nu_e$ contamination in the beam.
Finally, \minerva{} will search for the analogous neutral
current process $\nu p\to\nu p$ which is experimentally more
challenging due to backgrounds from non-leptonic processes.

The study described in this paper is a preliminary comparison of muon
kinematics in CCQE candidates between data and simulation with a
sample of $0.4\times10^{20}$ POT in the LE anti-neutrino beam.  These
data were taken during the construction of the \minerva{} detector when
only approximately one half of the detector had been installed.  The
scintillator tracker fiducial mass for this sample is $3.0$ of the
$6.4$~metric ton fiducial mass available in the full detector.  This sample is
$12\%$ of the total exposure measured as the product of POT and
fiducial mass.  The reference simulation is GENIE 2.6.2~\cite{GENIE}.
The CCQE cross-section at nucleon level comes from the derivation of
 Llewellyn-Smith~\cite{LS-CCQE} with vector form factors from the
BBBA2005 parametrization~\cite{BBBA2005}, the pseudo-scalar form
factor from PCAC, and the axial form factor in the dipole form with
$m_A=0.99$~GeV/c$^2$.  The nuclear effects when scattering from carbon
come from the Fermi gas model of Bodek and
Ritchie~\cite{Bodek-Ritchie}, and Pauli blocking is implemented by a
requirement that the outgoing nucleon has momentum of above the Fermi
momentum of $221$~MeV/c.  Final state interactions of outgoing hadrons
from the hard scattering process is simulated by INTRANUKE~\cite{GENIE}.

\section{Selection of CCQE Candidates}

There are many possible approaches to the selection of CCQE candidates
in the \minerva{} detector.  We expect, eventually, to have
at least three strategies for selection, each of which will select different
CCQE-rich final states and admit different backgrounds in the
analysis.  First, we can perform a topological track based analysis with a
one-track (muon) low $Q^2$ component and a two-track (muon and
nucleon) component at higher $Q^2$ where reconstruction of the
nucleon, at least when it is a proton, is feasible.  This style of
analysis can be most easily compared to results using similar
techniques from
SciBooNE~\cite{SciBooNE-CCQE} and NOMAD~\cite{NOMAD-CCQE}.  Second, we can
perform an inclusive analysis that selects only a muon and removes
background based on a veto on Michel electrons from the chain
$\pi^+\to\mu^+\nu_\mu\to e^+\nu_e\bar{\nu}_\mu\nu_\mu$ which is most similar to the
analysis from the MiniBooNE~\cite{MiniBooNE-CCQE-1,MiniBooNE-CCQE-2} 
experiment.  Finally,
one can identify tracks as in the first analysis but remove events
with additional visible recoil energy in the detector which is far
from the vertex.  The latter stipulation is important for keeping
events with low energy nucleons in the final state from the remaining
nucleus after the CCQE interaction.  The analysis presented here as
our first preliminary work towards this reaction is in the final of
these three styles.  As we progress in this analysis, we plan to add the 
other techniques.

\begin{figure}
\hspace*{0.05\textwidth}
\includegraphics[width=0.4\textwidth]{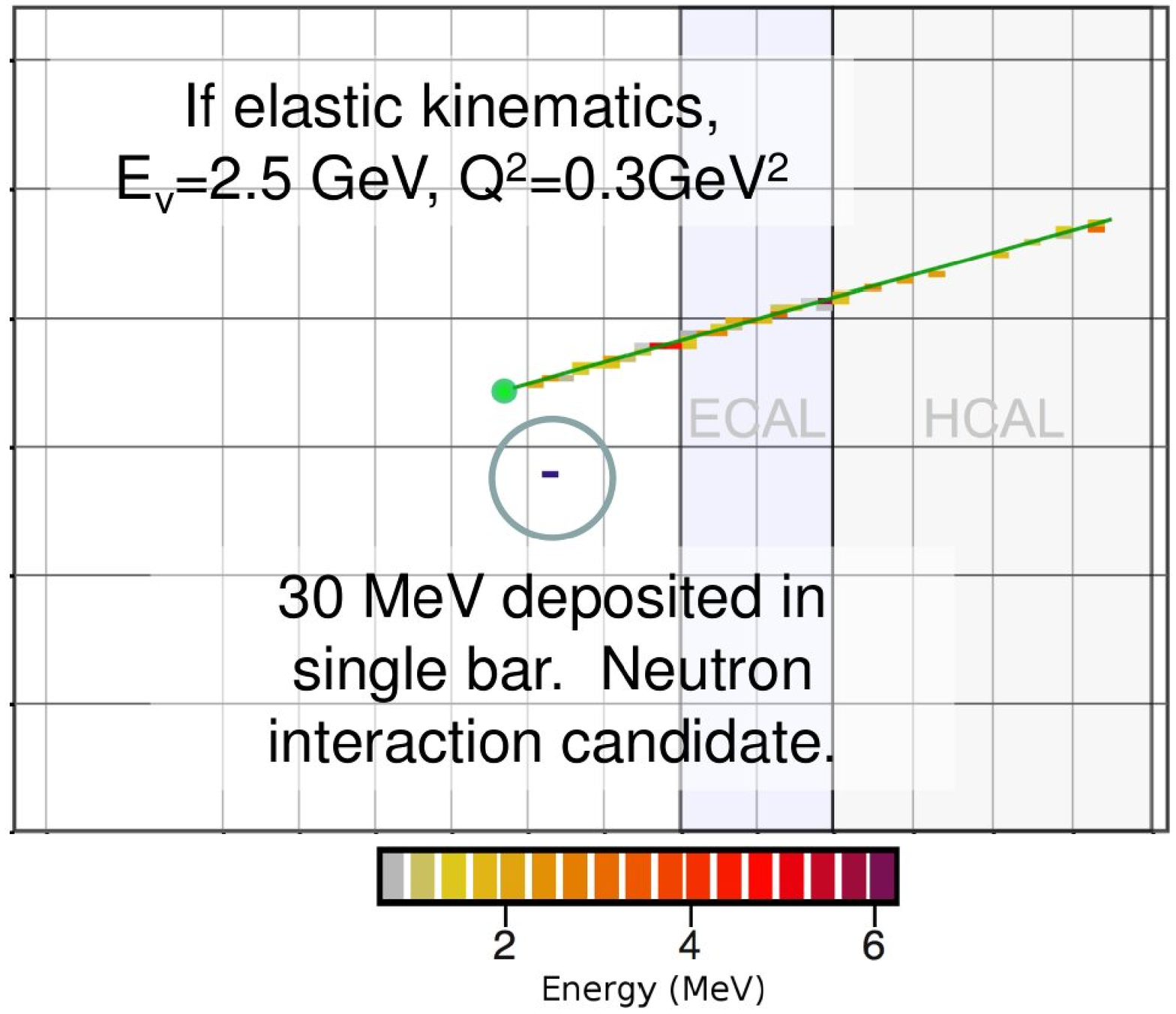} 
\hspace*{0.1\textwidth}
\includegraphics[width=0.4\textwidth]{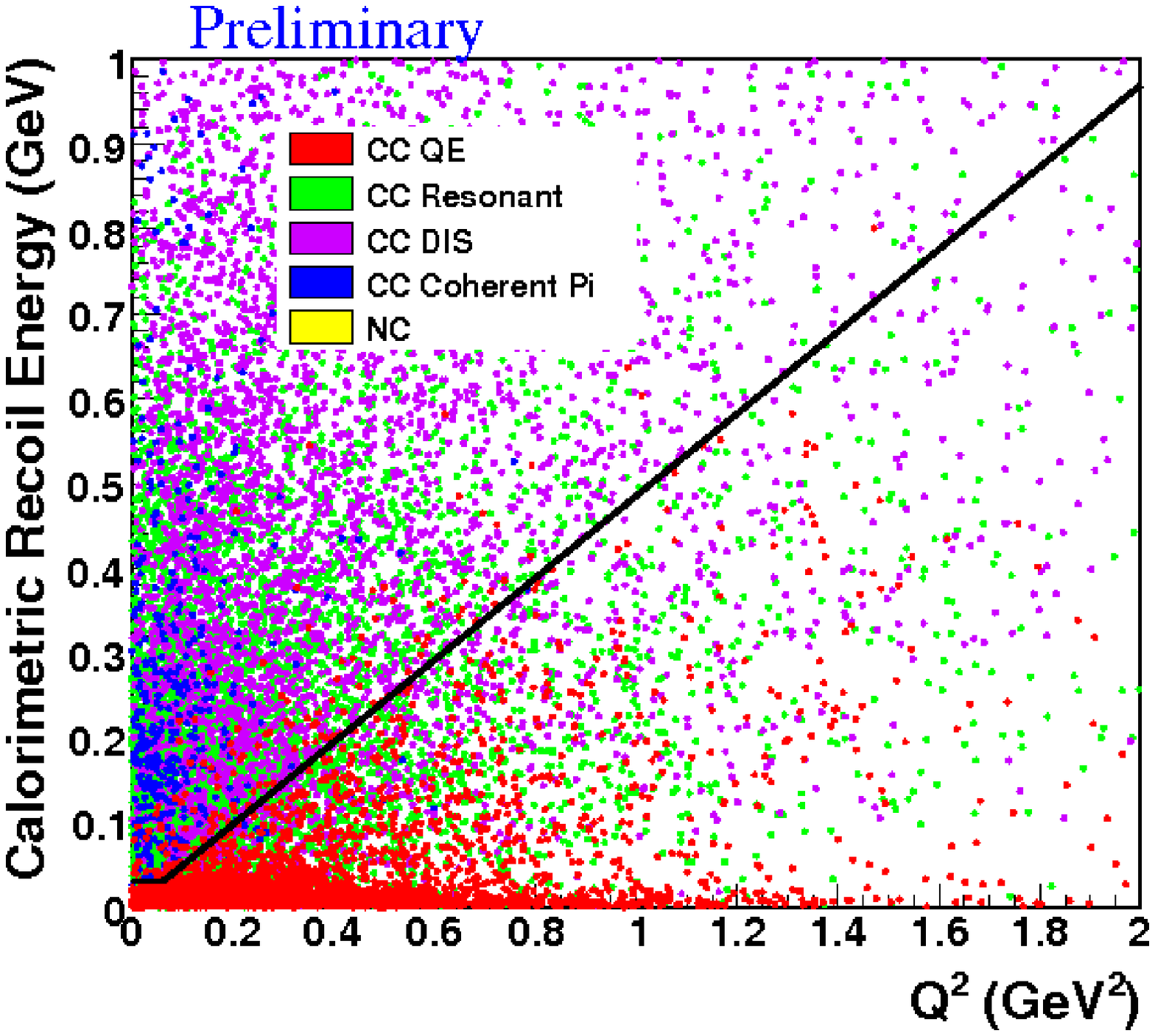} 
\caption{(left) A candidate event in the search for $\bar{\nu}_\mu
p\to\mu^+n$.  The track originates in the scintillator region and is
  momentum analyzed in the MINOS near detector.  The $30$~MeV energy
  deposit away from the track is consistent with the direction of 
  the recoil neutron.  (right) For simulated events, the recoil 
  energy away from the vertex 
  vs. $Q^2$ derived from the muon energy and angle.  The region
  selected in this space keeps very high efficiency but lets in
  significant background at high $Q^2$.}
  \label{fig:candidate}
\end{figure}

This analysis first selects events with a 
$\mu^+$ originating from the scintillator
tracker whose momentum and charge are measured in the MINOS near
detector~\cite{nuint11-schmitz}.  From the $\mu^+$ reconstructed
momentum and angle with respect to the neutrino beam direction we can
reconstruct the neutrino energy and $Q^2$,
\begin{eqnarray}
E_\nu^{\rm\textstyle rec}
&=&
\frac{m_n^2+2E_\mu\left(m_p-E_B\right)-\left(m_p-E_B\right)^2-m_\mu^2}{2 \left(m_p-E_B-E_\mu+\cos\theta_\mu\sqrt{E_\mu^2-m_\mu^2}\right)} \label{eqn:enu}\\
{\rm\textstyle and}~(Q^2)^{\rm\textstyle rec}
&=&
2E_\nu^{\rm\textstyle
rec}\left(E_\mu-\cos\theta_\mu\sqrt{E_\mu^2-m_\mu^2}\right)-m_\mu^2, \label{eqn:qsq}
\end{eqnarray}
where $m_n$, $m_p$ and $m_\mu$ are the neutron, proton and muon
masses, $E_\mu$ and $\theta_\mu$ are the muon energy and angle, and
$E_B$ is the target proton binding energy, $34$~MeV in our estimation
of the neutrino kinematics.

A quasi-elastic candidate event
passing this selection is shown in Fig.~\ref{fig:candidate}.  The
only other visible energy in the event is the single bar with a
$30$~MeV deposit consistent with the expected direction of recoil based
on the direction and energy of the muon.
The expected recoil is a neutron with kinetic energy of
$150$~MeV.  Typically only a small fraction of the neutron energy is
observed calorimetrically in the \minerva{} detector.

\begin{figure}
\includegraphics[width=0.33\textwidth]{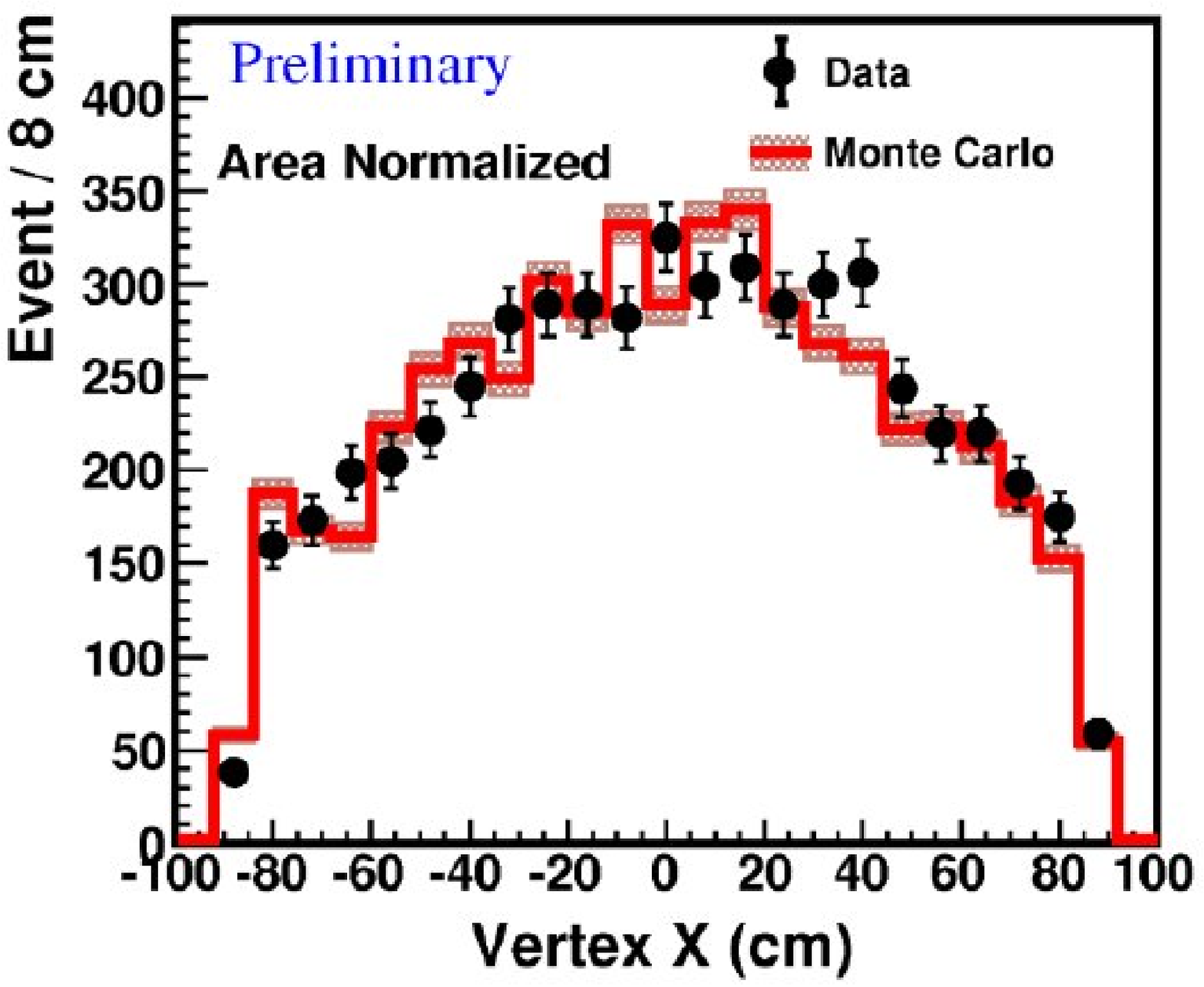} 
\includegraphics[width=0.33\textwidth]{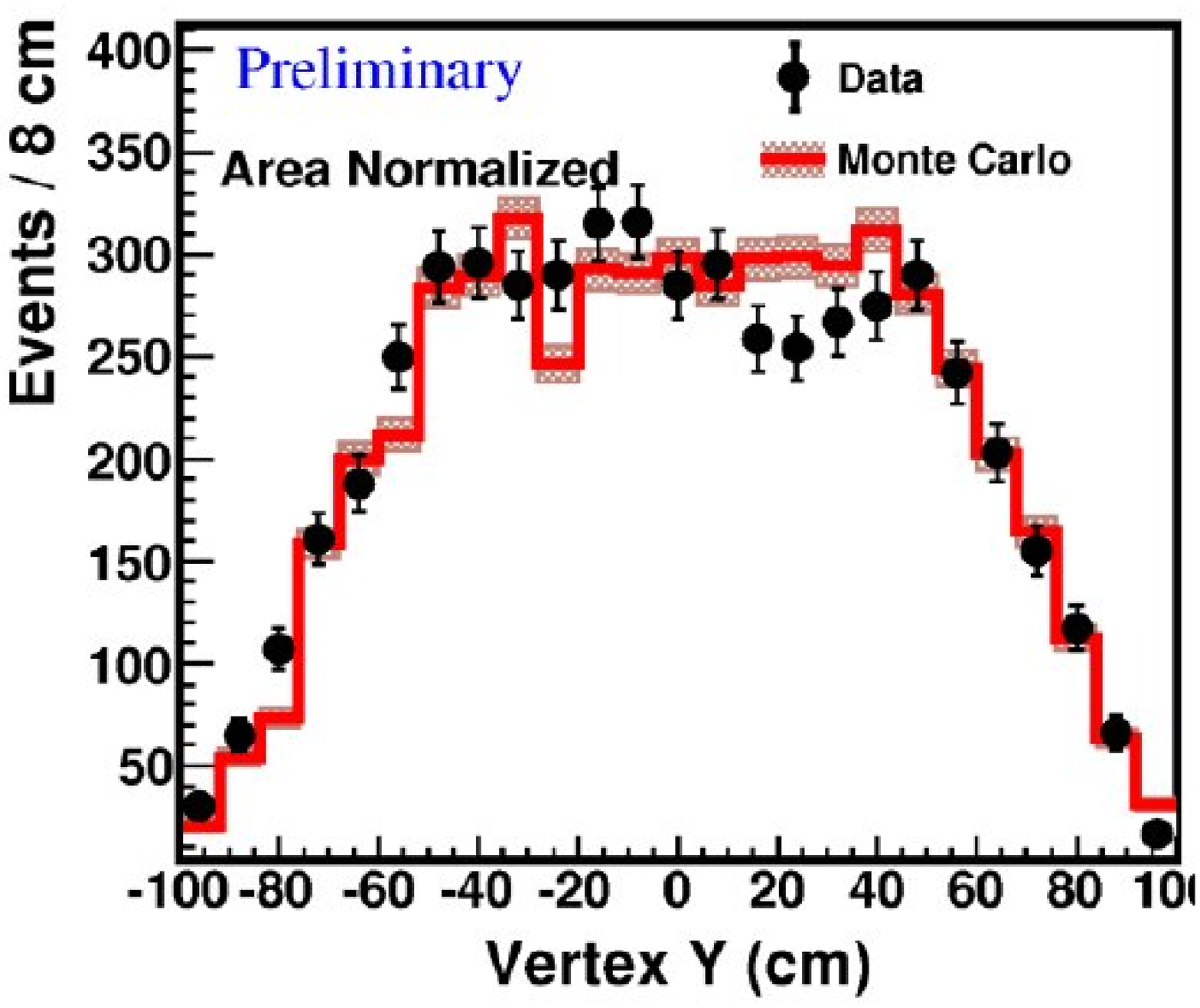} 
\includegraphics[width=0.33\textwidth]{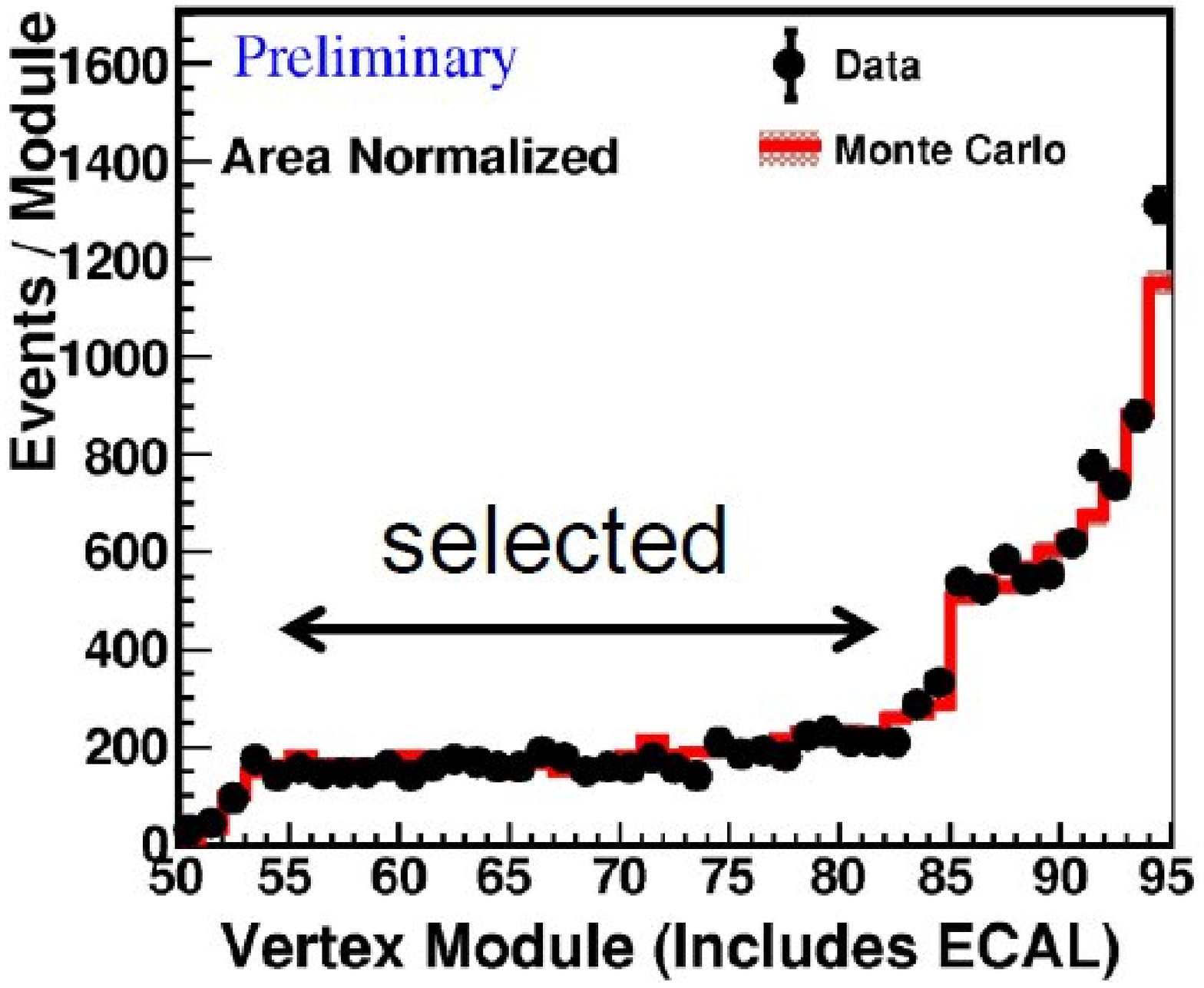}
\caption{Distributions of the position at which the muon in $\bar{\nu}_\mu
p\to\mu^+n$ originates.  (left and middle) The transverse directions,
  $x$ and $y$ within the hexagonal fiducial volume.  (right) The
  direction along the beam, including regions outside the fiducial
  volume selected by the analysis such as the more dense
  electromagnetic calorimeter (ECAL) region downstream of the fiducial
  volume and the very dense hadron calorimeter (HCAL) farther
  downstream.}  
\label{fig:positions}
\end{figure}

For this first analysis, we chose a very conservative cut on the
recoil energy in order to keep the efficiency of the analysis high.
This approach, however, lets in significant backgrounds, particularly
at high $Q^2$ as shown in Fig.~\ref{fig:candidate}.  For events
originating within the fiducial volume, the asymptotic efficiency at
high energies is $\approx 65\%$. The efficiency is highest at low
$Q^2$ and falls slowly to $\approx 30\%$ at high $Q^2$ not because of
the low recoil selection, but because of the decrease in muon
acceptance to enter the MINOS near detector. 
A hit-level GEANT4-based simulation of the detector
is used to evaluate these efficiencies and correct the neutrino
interaction simulation for detector efficiencies and resolutions.  We
gain confidence in this simulation by comparison of distributions of
candidate events in the data with the simulation in quantities that
are largely independent of the neutrino flux and the interactions.
For example, Fig.~\ref{fig:positions} shows that the distribution of
candidate vertices in the transverse and longitudinal dimensions of
the detector agrees with a relatively normalized simulation.

\section{Results}

We present the results as comparisons to the simulation, including
background events generated as described above.  The simulation also
includes our prediction of the untuned flux from a beamline simulation
which has large uncertainties that we expect to reduce in the future~\cite{nuint11-jerkins}.  
We evaluate these uncertainties in the {\em a priori} neutrino flux to
be lowest, about $7\%$, at the peak energy of the flux shown in
Fig.~\ref{fig:spectrumRates} and significantly larger above this
``focusing peak'' energy with an asymptotic uncertainty at high energy
of $16\%$.  These large flux uncertainties must be included in the
comparisons between data candidates and the simulation.  These
uncertainties allow a significant distortion in shape of the neutrino
flux with neutrino energy, $E_\nu$.  However, since the $Q^2$ spectrum
as a function of
$E_\nu$ is approximately independent of $E_\nu$, at least far below the
kinematic limit, $Q^2\stackrel{<}{\sim}2m_pE_\nu$, we do not expect
the flux uncertainties to significantly distort the predicted $Q^2$
distribution.

\begin{figure}
\includegraphics[width=0.49\textwidth]{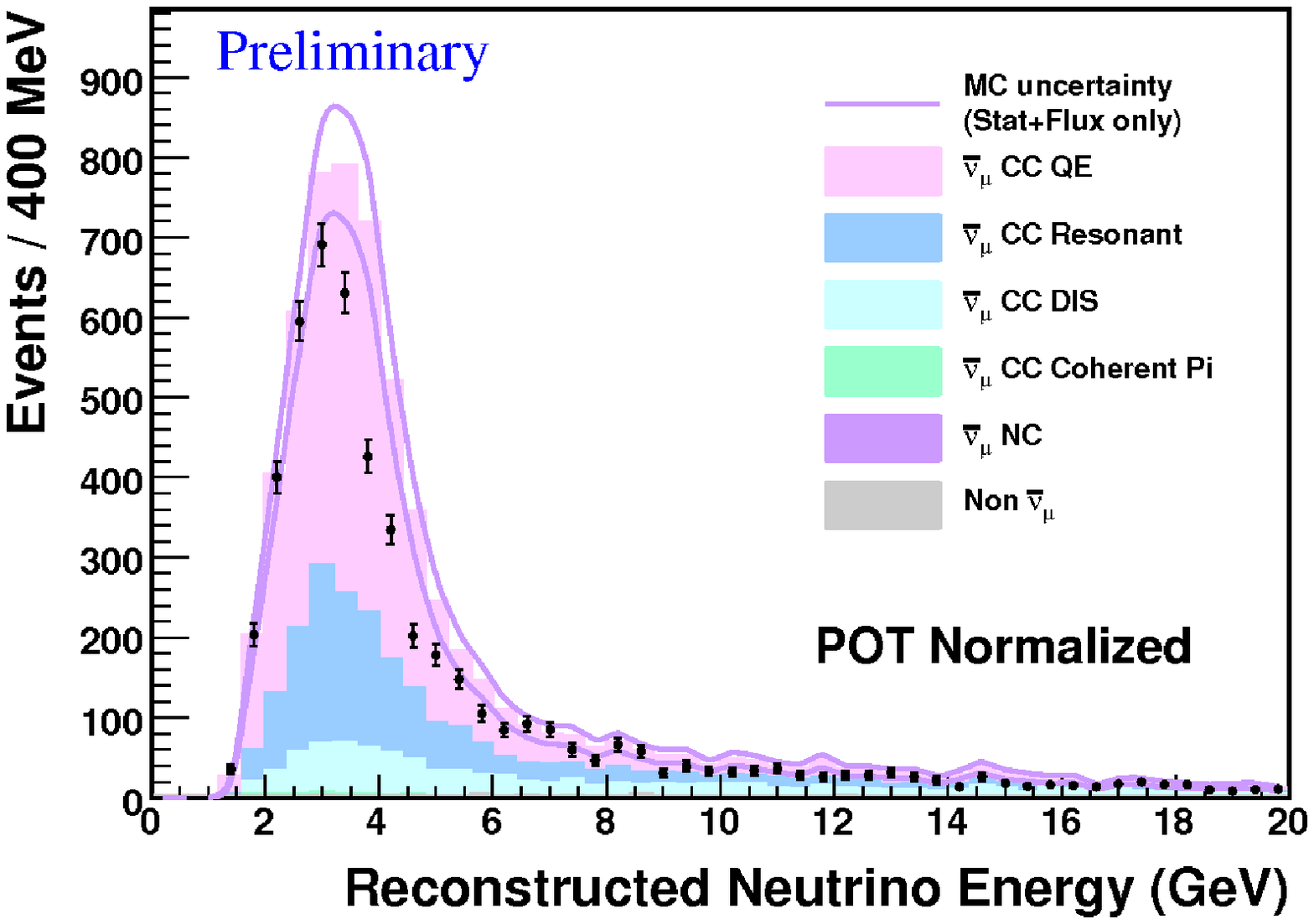} 
\includegraphics[width=0.49\textwidth]{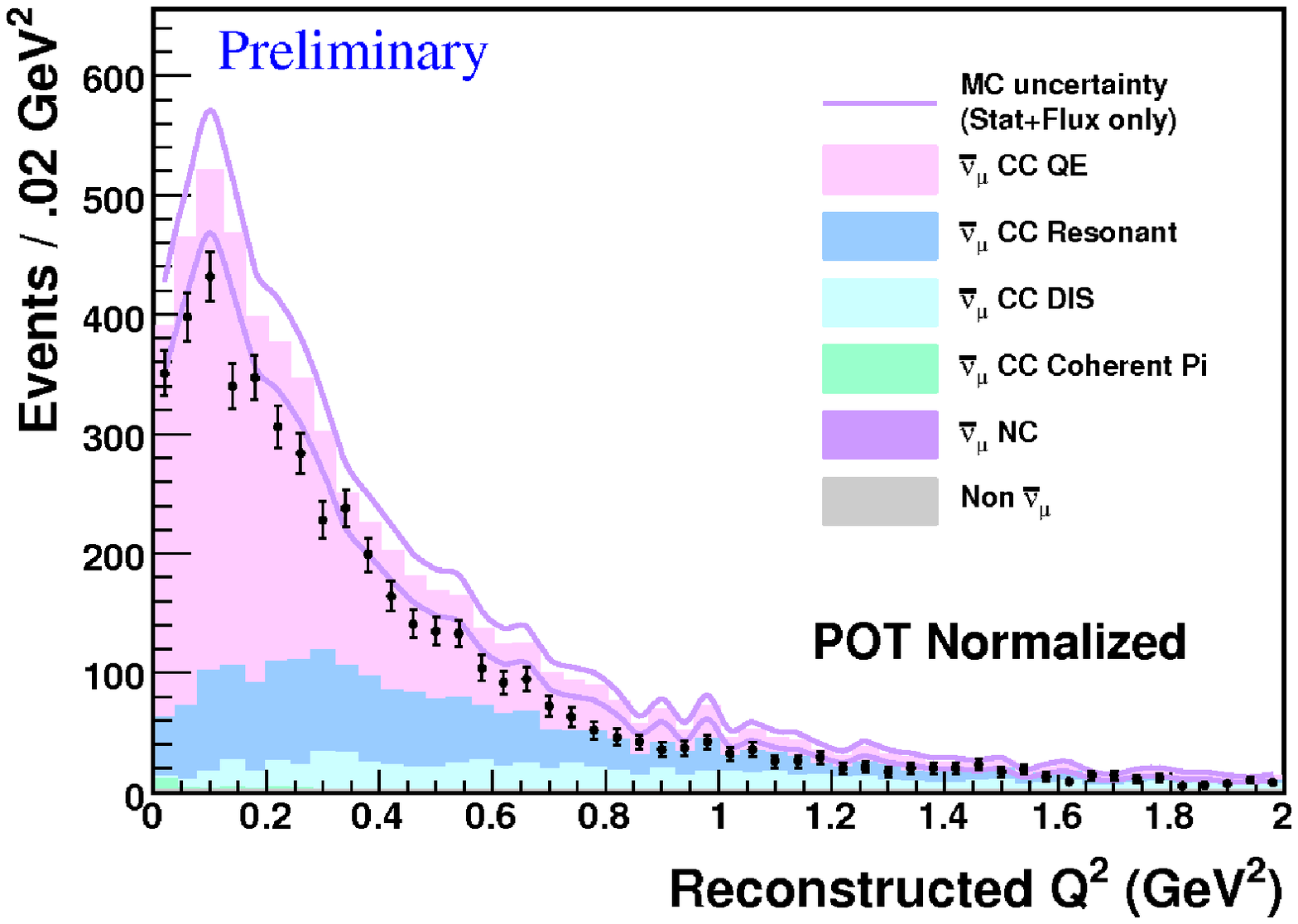} 
\caption{$E_\nu$ (left) and $Q^2$ (right) reconstructed from muon
kinematics by Eqns.~\ref{eqn:enu} and \ref{eqn:qsq}, respectively, for 
$\bar{\nu}_\mu p\to\mu^+ n$ candidates.
Different reactions in the simulated event sample are
shown stacked in the solid histogram.  The
flux uncertainty is shown as a band around the simulation prediction.}  
\label{fig:enu-qsq}
\end{figure}
\begin{figure}
\includegraphics[width=0.9\textwidth]{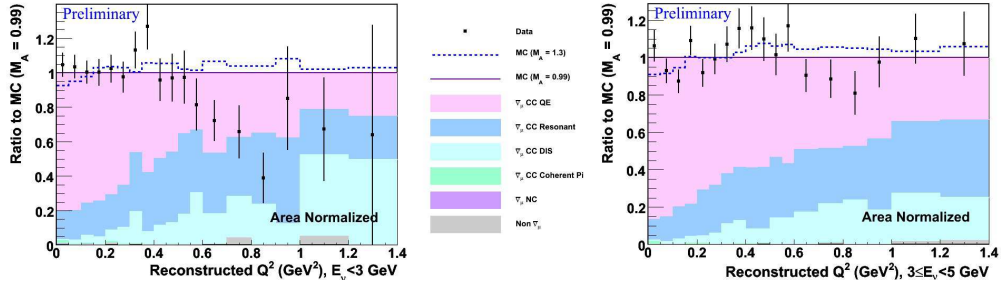} 
\caption{The relatively normalized ratio of data to simulated $Q^2$
  distributions for $E_\nu<3$~GeV (left) and $3<E_\nu<5$~GeV (right).
  The uncertainties shown are statistical only; flux uncertainties
  approximately cancel in the analysis of the shape of this
  distribution, detector systematics and background uncertainties
  which are not included in the shown uncertainties may not.  The
  change in shape expected if $m_A=1.3$~GeV is also shown.}
\label{fig:qsq-by-energy}
\end{figure}

Figure~\ref{fig:enu-qsq} shows the reconstructed neutrino energy and
$Q^2$ of the reactions for candidate events. The comparison shows a
disagreement outside the uncertainty band in the distribution of
neutrino energies just above the focusing peak, $3$~GeV$<E_\nu<6$~GeV.
However, the shape of the $Q^2$ distribution, at least in the region
where CCQE events make up the majority of the signal, is in reasonable
agreement with our simulation.  Note that the simulation uses an axial
mass of $0.99$~GeV.  MiniBoonE and SciBooNE, measuring quasi-elastic
neutrino and anti-neutrino scattering for $0.3<E_\nu<2$~GeV, prefer
significantly higher axial masses, consistent with
$m_A=1.3$~GeV~\cite{MiniBooNE-CCQE-1,MiniBooNE-CCQE-2,SciBooNE-CCQE},
whereas the NOMAD experiment, measuring quasi-elastic neutrino
scattering, $3<E_\nu<100$~GeV, prefers $m_A$ consistent with the value
of $0.99$~GeV assumed in our simulation.
Figure~\ref{fig:qsq-by-energy} shows the ratio of data to simulation
for the \minerva{} analysis in the low and high neutrino energy
portions of the sample.  Both appear more consistent with the low
$m_A$ assumed rather than the high $m_A$ of MiniBooNE and SciBooNE;
however, caution is warranted in drawing this conclusion because the
\minerva{} result only includes systematic
uncertainties from the neutrino flux, and does not yet include
uncertainties on the detector reconstruction or 
the backgrounds to the quasi-elastic interactions.

\section{Conclusions}

The \minerva{} collaboration has performed its first comparison of a
quasi-elastic rich anti-neutrino sample with expectations for a
range of assumed axial masses.  The
selection requires a muon from a neutrino interaction in the
scintillator target and a low observed recoil away from the vertex in
the final state to enhance elastic events.  \minerva{} possesses an
anti-neutrino sample approximately eight times as large as that in
this analysis on tape, and will be able to improve the background
rejection of this analysis in the future through a variety of
techniques that the capabilities of the detector make possible.

%%%%%%%%%%%%%%%%%%%%%%%%%%%%%%%%%%%%%%%%%%%%%%%%
%% BACKMATTER
%%%%%%%%%%%%%%%%%%%%%%%%%%%%%%%%%%%%%%%%%%%%%%%%

\begin{theacknowledgments}
This work was supported by the Fermi National Accelerator Laboratory,
which is operated by the Fermi Research Alliance, LLC, under contract
No. DE-AC02-07CH11359, including the MINERvA construction project,
with the United States Department of Energy. Construction support also
was granted by the United States National Science Foundation under NSF
Award PHY-0619727 and by the University of Rochester.  Support for
participating scientists was provided by NSF and DOE (USA) by
CAPES and CNPq (Brazil), by CoNaCyT (Mexico), by CONICYT (Chile), by
CONCYTEC, DGI-PUCP and IDI-UNI (Peru), and by Latin American Center
for Physics (CLAF). Additional support came from Jeffress Memorial
Trust (MK), and Research Corporation (EM).  Finally, the authors are
grateful to the staff of Fermilab for their contributions to this
effort and to the MINOS collaboration for their efforts to operate the
MINOS near detector and willingness to share its data.
\end{theacknowledgments}

%%%%%%%%%%%%%%%%%%%%%%%%%%%%%%%%%%%%%%%%%%%
%% The following lines show an example how to produce a bibliography
%% without the help of the BibTeX program. This could be used instead
%% of the above.
%%%%%%%%%%%%%%%%%%%%%%%%%%%%%%%%%%%%%%%%%%%

\bibliographystyle{aipproc}

\begin{thebibliography}{99}

\bibitem{nuint11-schmitz}
D.~Schmitz, ``The \minerva{} Neutrino Scattering Experiment'', these
proceedings (2011).

\bibitem{K2K-CCQE-1} R. Gran {\em et al.} [K2K Collaboration], Phys.\ Rev.\ {\bf D74},
052002 (2006).

\bibitem{K2K-CCQE-2} X. Espinal and F. Sanchez [K2K Collaboration], AIP
Conf.\ Proc.\ 967, 117 (2007).

\bibitem{MiniBooNE-CCQE-1} A. A. Aguilar-Arevalo {\em et al.} [MiniBooNE Collaboration], Phys.\ Rev.\ Lett.\ {\bf 100}, 032301 (2008)

\bibitem{MiniBooNE-CCQE-2} A. A. Aguilar-Arevalo {\em et al.} [MiniBooNE Collaboration], Phys.\ Rev.\ {\bf D81} 092005 (2010).

\bibitem{SciBooNE-CCQE} J. L. Alcaraz-Aunion and J. Walding [SciBooNE Collaboration], AIP Conf.\ Proc.\ 1189, 145 (2009).

\bibitem{NOMAD-CCQE} V. Lyubushkin {\em et al.} [NOMAD Collaboration], Eur.\
Phys.\ Jour.\ {\bf C63}, 355 (2009).

\bibitem{GENIE} C. Andreopoulos {\em et al.}, Nucl.\ Instrum.\ Meth.\ {\bf A614} 87 (2010).

\bibitem{LS-CCQE} C. H. Llewellyn Smith, Phys.\ Rept.\ {\bf 3} 261 (1972).

\bibitem{BBBA2005} A.~Bodek, R.~Bradford, H.~Budd and J.~Arrington,
Nucl.\ Phys.\ Proc.\ Suppl.\ {\bf 159} 127 (2006).

\bibitem{Bodek-Ritchie} A.~Bodek and J.~Ritchie, Phys.\ Rev.\
{\bf D23} 1070 (1981).

\bibitem{nuint11-jerkins}
M.~Jerkins, ``Measuring the NuMI Beam Flux for \minerva{}'', these proceedings (2011).

\end{thebibliography}

\end{document}